\documentstyle[preprint,eqsecnum,aps]{revtex}
\tightenlines
\newcommand{\be}{\begin{equation}}
\newcommand{\ee}{\end{equation}}
\newcommand{\bea}{\begin{eqnarray}}
\newcommand{\eea}{\end{eqnarray}}

\begin{document}
\draft
\title{ 
{Extended Self-Similarity in the Two-Dimensional Metal-Insulator Transition}}
\author{L. Moriconi}
\address{Instituto de F\'\i sica, Universidade Federal do Rio de Janeiro,\\
C.P. 68528, Rio de Janeiro, RJ -- 21945-970, Brasil}
\maketitle
\begin{abstract}
We show that extended self-similarity, a scaling phenomenon firstly observed in classical turbulent flows, holds for a
two-dimensional metal-insulator transition that belongs to the universality class of random Dirac fermions. Deviations from multifractality, which in turbulence are due to the dominance of diffusive processes at small scales, appear in the condensed matter context as a large scale, finite size effect related to the imposition of an infra-red cutoff in the field theory formulation. We propose a phenomenological interpretation of extended self-similarity in the metal-insulator transition within the framework of the random $\beta$-model description of multifractal sets. As a natural step our discussion is bridged to the analysis of strange attractors, where crossovers between multifractal and non-multifractal regimes are found and extended self-similarity turns to be verified as well.
\end{abstract}
\vfill
\narrowtext

\section{Introduction}
The surprising observation of extended self-similarity (ESS) in fully developed turbulence \cite{benzi}, a hidden scaling behavior which takes place in the crossover between inertial and dissipative ranges, has provided an empirical -- and very efficient -- way of determining accurate scaling exponents of velocity structure functions. The essential assumption of the method is that there are no anomalous corrections to Kolmogorov's $4/5$ law, as required from the existence of a constant energy flux through the inertial range \cite{frisch}. To state in an explicit way what is meant by ESS, let the average
\be
S_q(r) = \langle |\vec v (\vec x_1) - \vec v (\vec x_2)|^{q} \rangle \equiv \lim_{T \rightarrow \infty} 
{1 \over {T}} \int_0^T dt |\vec v (\vec x_1,t) - \vec v (\vec x_2,t)|^{q}
\label{i1}
\ee
denote the $q$-th order structure function, where $r=|\vec x_1 -\vec x_2|$ and $\vec v ( \vec x,t )$ is the velocity field. In the inertial range, the scaling result $S_q(r) \sim r^{\zeta (q)}$ holds. At smaller length scales, where dissipation effects come into play, $S_q(r)$ loses its power law form. However, a plot of $G_q(r)=\ln [S_q(r)]$ versus $G_p(r)=\ln [S_p(r)]$, shows in a clear way that the linear relation 
\be
G_p(r)={{\zeta(p)} \over {\zeta(q)}} G_q(r) + c(p,q) \label{i2}
\ee
is reasonably verified even in the crossover towards the dissipative scale. Since Kolmogorov's $4/5$ law predicts $\zeta(3)=1$, improved evaluations from experimental data can be obtained for $\zeta(p)$, using $q=3$ above. 
Later on, it was realized that a refinement of ESS, referred to as ``Generalized Extended Self-Similarity'' would lead to better linear fits \cite{benzi2,bif}. Actually, one finds that if $G_n(r)$ is taken as fixed ``reference structure function", for some arbitrary $n$, then $G_{q/n}(r) \equiv G_q(r)-(q/n)G_n(r)$ satisfies to a higher degree of precision a relation similar to (\ref{i2}), with slope given now by $[n \zeta(p) - p \zeta(n)]/[n \zeta(q)-q\zeta(n)]$. From now on we will specify, only when necessary, the former definition of ESS, by ``ESS-I" while its generalized version will be named ``ESS-II".

It is clear ESS-I leads to ESS-II, but the converse may be not true. The definition of ESS-II is indeed necessary to cope with the fact that at the smallest scales one expects to have $S_q(r) \sim r^q$. It is not difficult to prove that 
ESS-II implies structure functions may be written as 
\be
S_q(r)=c_q [f_1(r)]^{q}[ f_2(r) ]^{\zeta(q)} \ , \ \label{i3}
\ee
where $f_1(r) \rightarrow 1$, $f_2(r) \sim r$ in the inertial range and $f_1(r) \sim r$, $f_2(r) \rightarrow 1$ when $r \rightarrow 0$. At present, there is no theoretical understanding of the simple crossover behavior (\ref{i3}) for the structure functions in a turbulent fluid. We  note, however, that in the slightly different context of the turbulent transport of a passive scalar, an analytical proof of ESS-I was carried out in ref. \cite{segel}.

The idea that ESS could hold for a larger class of models is supported not only by the problem of the transport of a passive scalar, but also by numerical investigations of kinetic roughening \cite{puny} and magnetohydrodynamic turbulence \cite{basu}. It is tempting to look for ESS whenever multifractal scaling laws take place, adventuring beyond the turbulence arena where it has been observed. This is precisely the task we pursue in this work for a few additional models. Our motivation is twofold. We stress that the phenomenological description of multifractality in turbulence, based on the random $\beta$-model approach to the Richardson cascade \cite{benzi3}, suggests that ESS could be analogously found in quantum problems like condensed-matter localization (see sec. II). Furthermore, we have in mind the conjecture that several multifractal systems could be mapped into specific patterns, likely through a field theory point of view \cite{dupla}. Here, inspiration comes from the sucessful applications of field theoretical methods in critical phenomena, where universality classes and critical properties of statistical systems at thermodynamic equilibrium proved suitable for systematic study. In analogy with the renormalization group strategy, firmly based on the experiment, we are now faced, thus, with the primary problem of recognizing, among the plurality of multifractal models, common features -- like ESS -- that could be derived from some general theory. Of course, an important role in this scenario is to be played by a number of well-established multifractal field theory models, as the ones for random spin systems \cite{lud2}, Anderson localization \cite{peliti}, the two-dimensional metal-insulator transition \cite{lud3,chalker,cast}, hadron  jets in QCD \cite{brax}, etc.

While studying some completely diverse models, we have structured the paper to be as self-contained as possible. In the next section we discuss a two-dimensional metal-insulator transition, where we respectively verify ESS-I and ESS-II for strong and weak disordered regimes of the cutoff field theory. We study the transition via its modeling in terms of two-dimensional Dirac fermions \cite{lud3,chalker,cast}. A ``cascade description" of the metal-insulator transition at the critical point is addressed, which takes us naturally to the analysis of strange attractors, the subject of sec. III. We study then the two perhaps most celebrated strange attractors, H\'enon \cite{henon} and Lorenz \cite{lorenz}, collecting clear evidence for the existence of ESS-II and ESS-I, respectively, in these geometrical sets. In sec. IV, we put forward an analysis of ESS in terms of the random $\beta$-model description of multifractality, which, albeit essentially phenomenological, throws some light on the previous findings. In sec. V, we comment on our results and point directions of further research.

\section{The Two-Dimensional Metal-Insulator Transition}
The multifractal nature of delocalized states at the metal-insulator transition in condensed matter systems has
been attracting a large deal of interest in recent years \cite{jansen}. We propose here a qualitative cascade picture to approach the physics of delocalized multifractal wavefunctions. A more quantitative, although phenomenological, description of the ``cascade picture" of the metal-insulator transition will be postponed to sec. IV.

Suppose that at initial time a smooth and localized wavepacket is defined, centered around an arbitrary point $P$. As the system evolves, quantum propagation in the disordered background will make the wavepacket to spread and fragment, so that at very large times and very far from $P$, a cascade for the probability density develops, yielding a multifractal measure. Of course, multifractality is lost at scales of the order of the system's size, where the smoothness of the wavefunction is recovered due to reflections at the boundaries. Such a crossover between two scaling regimes is expected to be found in any experiment or realistic model of electron localization. The above state of affairs is analogous to the one encountered in turbulence, if the roles of infrared and ultraviolet regions are exchanged. Eddy fragmentation in turbulence gives rise to an energy cascade, and dissipation at smaller scales, where viscous effects are more important than convection, render the velocity configurations smooth.

Our prototype for the study of the multifractal-nonmultifractal crossover of probability density functions will be an exactly solvable model which describes a two-dimensional metal-insulator transition, closely related to the one observed in the integer quantum Hall effect \cite{lud3,chalker,cast}. The model consists of Dirac fermions moving in a box of size $L \times L$ under the presence of a transverse random magnetic field. Deviations from multifractality follow from the imposition of periodic boundary conditions in the $x$ and $y$ directions, introducing an infra-red cutoff in the corresponding field theory model. The random Dirac hamiltonian is given by
\be
H= \sum_{\mu=1,2} \sigma_\mu [ i \partial_\mu - A_\mu ] \ ,\ \label{mit1}
\ee
where $A_\mu$ is a random gauge field and the $\sigma_\mu$'s are Pauli matrices.

The exact multifractal spectrum computed by Castillo at al. \cite{cast} is obtained from the zero modes of (\ref{mit1}).  The normalized solution of $H \psi = 0$ is just
\be
\psi_0(\vec x)= {\cal N}^{-1} \exp[ - \phi(\vec x) \sigma_3 -i \chi (\vec x) ]
\left[
\matrix{ 1\cr
         0} \right] \ , \ \label{mit2}
\ee
where the gauge field \footnote{Note that the total magnetic flux on the two-dimensional space is zero.
Exact results were obtained so far only in the trivial topological sector.} has been written as
$A_\mu (\vec x)= \sum_\nu \epsilon_{\mu \nu} \partial_\nu \phi (\vec x) + \partial_\mu \chi (\vec x)$ and
\be
{\cal N}=\left [ \int d^2 \vec x \exp ( -2 \phi(\vec x))
\right ]^{1 \over 2} \ . \ \label{mit3}
\ee
The random magnetic field $B(\vec x)=- \nabla^2 \phi(\vec x)$ is logarithmically correlated, that is, its 
fluctuations are described by the probability density functional
\be
P[ \phi ] \propto  \exp [ - { 1 \over {2g}} \int d^2 \vec x ( \vec \nabla \phi )^2  ]  \ . \ \label{mit4}
\ee

We are interested to study amplitude fluctuations of $\psi_0(\vec x)$, which do not depend on the phase $\chi(\vec x)$. 
Let $\Omega$ be a square box of size $r \times r$. We define now structure functions of order $q$ as
\be
S_q(r)=\langle \left [ \int_{\Omega} | \psi_0|^2 d^2
\vec x \right ]^q  \rangle \ . \ \label{mit5}
\ee
The above average is computed in a quenched disorder scheme from the probability density functional (\ref{mit4}). According to the exact results \cite{cast} we have $S_q(r) \sim r^{(\tau(q)+2)}$ for $r \ll L$, where $\tau(q)$ is the scaling exponent given in table 1. Strong and weak regimes are defined, respectively, by $q_c \leq 1$ and $q_c > 1$, where $q_c =\sqrt{2 \pi / g}$.
\vspace{0.2cm}
\begin{center}
\begin{tabular}{|c|c|c|} \hline
~ Structure Function Order ~ & ~ $\tau(q)$ (weak disorder) ~ & ~ $\tau(q)$ (strong disorder) ~ \\ \hline
$|q| \leq q_c$  & $ 2(q-1)(1-q/q_c^2)$ & $-2(1-q/q_c)^2$ \\ \hline
$|q| > q_c$  & $ 2q(1-{\hbox{sgn}}(q)/q_c)^2$ & $ 4(q-|q|)/q_c$  \\ \hline
\end{tabular}
\end{center}
\vspace{0.1cm}
{\centerline{\hbox{{\bf{Table 1.}} The scaling exponent $\tau(q)$ for the weak and strong disorder regimes.}}}
\vspace{0.2cm}

Computations were done through the Monte-Carlo method on a $600 \times 600$ lattice. The simulation consisted of $10^3$ Monte-Carlo steps per site. The averages were taken on a $ 100 \times 100$ sublattice. Weak and strong regimes were investigated, corresponding, respectively, to $q_c=10$ and $q_c=1$. Here, we report data for $q=3,6,9$ in the weak regime, and $q=-3,-6$ in the strong regime. We considered a single realization of $\phi(\vec x)$ for each pair $(q,q_c)$ of parameters. As a matter of fact, due to the self-averaging property of model (\ref{mit1}) \cite{cast}, a single realization of the multifractal wavefunction leads to scaling exponents identical to the ones computed from quenched averages in the large box limit. The numerical results are shown in Figs. 1 and 2, toghether with the exact slopes for comparison. We find in a very clear way that the weak regime is well-accounted for ESS-I, while the strong regime exhibits ESS-II. 

\section{Extended Self-Similarity in Strange Attractors}
There is an element in the cascade description of multifractal wavefunctions proposed above that brings our attention to the self-similar structure of strange attractors: the essential point is that the wavefunction normalization is preserved throughout the cascade. This simple constraint makes a bridge to the phenomenological analysis of strange attractors established by Benzi et al. \cite{benzi3}. In their work, a large and fixed number of $N$ points homogeneously distributed in some compact region of space (the basin of attraction) is sucessively iterated, producing an approximation to the attractor.
It is assumed a set with the same statistical properties would be generated after $N$ iterations of any arbitrary
initial point, for $N$ large enough. Structure functions of order $q$ may be defined as
\be
S_q(r)= \langle n(r)^q \rangle = {1 \over N} \sum_{i=1}^N \left [ \sum_{j=1}^N \Theta(d_{(i,j)}-r) \right ]^q \ , \ 
\label{sa2}
\ee
where $n(r)$ is the number of points contained in a ball of radius $r$, while $d_{(i,j)}$ is the distance between points labelled by $i$ and $j$, and $\Theta(x)$ is the Heaviside function. As $N \rightarrow \infty$, multifractality means that $S_q(r) \sim r^{\zeta (q)}$ at small scales. Since the total number of points $N$ is kept constant through iterations, the quantity $n(r)$ is immediately recognized as the analogous of $\int_\Omega d^2 \vec x |\psi|^2$ in the context of localization. The main difference between the dynamics of strange attractors and the evolution of delocalized wavefunctions, however, is the fact that the latter will not have a finite area support in general. 

The natural question we pose regards the corrections to the multifractal asymptotic behavior at large scales, comparable to the overall size of the attractor. We performed numerical experiments on the H\'enon and Lorenz strange attractors. The strange attractor generated by the H\'enon mapping \cite{henon},
\bea
&x_{n+1}&=1-ax_n^2+y_n \ , \ \nonumber \\
&y_{n+1}&=bx_n \ , \ \label{sa1}
\eea
with $a=1.4$ and $b=0.2$, is fully chaotic, containing no periodic orbits. As it is well-known, at certain range of scales the H\'enon attractor is locally like several approximately parallel stripes. At large enough scales, we may say, therefore, H\'enon attractor is roughly one-dimensional. However, sweeping a considerable larger range of scales, the fractal substructure shows up and one gets a non-trivial fractal dimension slightly greater than unity.
To study such a dimensional crossover, we have considered the H\'enon mapping for $10^5$ iterations, with initial point $x=y=0$. We focused our attention on the 2428 points contained in the region $|x-0.76|<0.1$, $|y-0.16| < 0.01$.
Defining now the functions $G_q(r) = \ln[S_q(r)]-q \ln r$, we clearly observe, from the results depicted in Fig. 3, for $q=3,6$ the existence of ESS-II in the H\'enon attractor. The plateaus observed at larger values of $r$ confirm the 
existence of one-dimensional large scale structures. 

Similar computations were carried out for the Lorenz attractor \cite{lorenz}. This dynamical system is given by the following coupled differential equations,
\bea
&\dot x& = -s(x-y) \ , \ \nonumber \\
&\dot y& = -xz+rx-y \ , \ \nonumber \\
&\dot z& = xy -bz \ , \ \label{sa3}
\eea 
where we take $s=10$, $b=8/3$ and $r=28$. The discrete flow $(x(t_n),y(t_n),z(t_n))$ which generates the strange attractor in the three-dimensional phase space is obtained using the time step $\delta = 0.01$ and $N=3000$. The initial point has coordinates $x=y=0$ and $z=0.01$. The first 2000 iterations were discarded. The range of scales investigated runs from the smallest ones to the overall size of the attractor, so that we expect now a crossover between a set with non-trivial scaling behavior and a set of vanishing dimension (the Lorenz attractor as ``viewed from the infinity"). We confirm ESS-I from Fig. 4, where the plots of $G_q(r)= \ln[S_q(r)]$ for $q=3$ and $q=8$ are shown. Note that $N$ is not a very large number, usually the condition to get good evaluations of the scaling exponents. The situation is here very similar to the one found in turbulence, when one extracts, through ESS, scaling exponents out of low Reynold's numbers data.

We have material evidence, therefore, to hypothesize that ESS is a property shared by other strange attractors and metal-insulator transitions. It is not clear, however, under which conditions ESS-I or strict ESS-II will hold.  Next, we attempt at a phenomenological description of ESS in general, under the light of random $\beta$-models.   

\section{Random $\beta$-model analysis}
The existence of ESS in multifractal systems seems to be intimately related to the dynamics of fragmentation. The random $\beta$--model approach to strange attractors and turbulence \cite{benzi3}, originally devised to capture the statistical properties associated with fragmentation events, is an interesting ground to get some clues on ESS.

The multifractal set given by a strange attractor is generated, in the random $\beta$-model line of thought, from the fragmentation of boxes in some $d$-dimensional phase-space of a dynamical system. At the $n$-th fragmentation step, a box with volume $\Omega_n=\ell_n^d$, containing a certain number of points $N_n$ is fragmented into sub-boxes with volumes 
$\Omega_{(n+1)}=\ell_{(n+1)}^d$, containing each of them $N_{(n+1)}$ points. The total volume of the ``newborn" boxes is a fraction $\beta_n$ (a random variable) of the original volume of the ``parent" box. In the random $\beta$-model we assume that at each step of fragmentation the new length scale is a fixed fraction of the previous one, that is, $\ell_{(n+1)} =\ell_n / a$, where $a>1$. Furthermore, the volume ratio $\beta_n$ varies from box to box, without any correlation, being described by some probability function $P(\beta)$. Since the total number of points is conserved, we may write
\be
\beta_n={ { N_n } \over { N_{(n+1)} } } \left ( { { \ell_{(n+1)} } \over { \ell_n } } \right )^d \ . \ \label{bm1}
\ee 
A simple relation may be obtained from (\ref{bm1}) for the number of points in a box generated in the
sequence of fragmentations $\beta_0 \rightarrow \beta_1 \rightarrow ... \rightarrow \beta_{(n-1)}$:
\be
N_n= \left( {  { \ell_n } \over { \ell_0 } } \right )^d \prod_{i=0}^{n-1} \beta_i^{-1} \ . \ \label{bm2}
\ee
Thus, we find that
\be
S_q( \ell_n) = \langle N^q_n \rangle = \left( { { \ell_n } \over { \ell_0 } } \right )^{\zeta(q)} \ , \ \label{bm3}
\ee
where 
\be
\zeta(q) = qd - \log_a \langle \beta^{-q} \rangle \ . \ \label{expnt}
\ee
Similar arguments may be advanced for the structure functions in turbulence, taking into account the energy cascade, giving $\zeta(q)=q/3-\log_a \langle \beta^{(1-q/3)} \rangle $.

Observe that for a given $a>1$, there is a discrete set of $a^d$ possible values of $\beta \leq 1$. 
In order to approach metal-insulator transitions, we introduce a variant of the random $\beta$-model which allows not only
a larger set of $\beta$'s, but also to have $\beta > 1$. In this way, statistical fluctuations of the random variable $\beta$ can be determined, in good approximation, from some probability density function $\rho(\beta)$ with $0 \leq \beta < \infty$. To define the alternative model, consider the union $U_0$ of $M_0$ boxes of volumes $\ell_0^d$, containing each $N_0$ points. We call $U_0$ a ``coherence region", since after the first fragmentation step, by imposition, a region $U_{01}$ is produced, the union of $M_{01}$ boxes of volumes $\ell_1^d$, each one taking the same number of $N_{01}=M_0 N_0/ M_{01}$ points. We have $\beta = M_{01}/M_0 a^{-d} =   N_0/N_{01} a^{-d}$, which may be now larger than unity. The model is supplemented by a ``splitting rule" (deterministic or not) that states if $U_{01}$ is still a coherence region or if it will be broken into a number $p$ of independent sub-coherence regions $U_{01(1)}$, $U_{01(2)}$, ..., $U_{01(p)}$, all of them having identical volumes. The same procedure is then inductively repeated as fragmentation proceeds. If, for instance, coherence regions are initially split into two sub-regions, we get the following diagram of fragmentations:
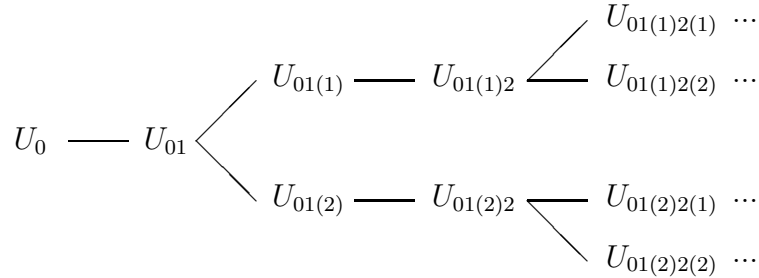
\begin{center}
\unitlength=1mm
\begin{picture}(60,40)(0,0)
\put(-18,20){\makebox(0,0){$U_0$}}
\put(-13,20){\line(1,0){8}}
\put(0,20){\makebox(0,0){$U_{01}$}}
\put(4,20){\line(1,1){8}}
\put(4,20){\line(1,-1){8}}
\put(19,28){\makebox(0,0){$U_{01(1)}$}}
\put(19,12){\makebox(0,0){$U_{01(2)}$}}
\put(25,28){\line(1,0){8}}
\put(25,12){\line(1,0){8}}
\put(41,28){\makebox(0,0){$U_{01(1)2}$}}
\put(41,12){\makebox(0,0){$U_{01(2)2}$}}
\put(48,28){\line(1,0){8}}
\put(48,12){\line(1,0){8}}
\put(48,28){\line(1,1){8}}
\put(48,12){\line(1,-1){8}}
\put(66,28){\makebox(0,0){$U_{01(1)2(2)}$}}
\put(66,12){\makebox(0,0){$U_{01(2)2(1)}$}}
\put(66,36){\makebox(0,0){$U_{01(1)2(1)}$}}
\put(66,4){\makebox(0,0){$U_{01(2)2(2)}$}}
\put(77,28){\makebox(0,0){...}}
\put(77,12){\makebox(0,0){...}}
\put(77,36){\makebox(0,0){...}}
\put(77,4){\makebox(0,0){...}}
\end{picture}
\end{center}
After $n$ iterations, a box of volume $\ell_n^d$ will contain a number of points that is still given by (\ref{bm2}). Since all paths in the fragmentation diagram have equivalent probabilistic weights, for reasonable choices of the splitting rule, then it follows that both (\ref{bm3}) and (\ref{expnt}) hold as well, if the number of coherence regions is large enough to allow for statistical averaging.

As an application of the above ideas, we identify $N_n/N$ with the box probability $\int_{\Omega_n} |\psi|^2$ in condensed matter localization. Recalling the discussion of sec. II, we reproduce the scaling exponents for the ranges $|q| \gg q_c$ in the weak and strong regimes of disorder in terms of a bifractal model. Let us take $\beta_- < 1 < \beta_+$ as the volume ratios, which are generated with probabilities $P_-$ and $P_+=1-P_-$, respectively. When $q \rightarrow \pm \infty$, it folows that $\langle \beta^{-q} \rangle \rightarrow P_\mp \beta_\mp^{-q}$. We find that both disorder regimes correspond to $P_+=P_-=1/2$ and $a=\sqrt{2}$. Hovewer, the weak disorder regime
gives
\be
\log_2 \beta_\pm =  {1 \over q_c} ( {1 \over q_c} \pm 2) \ , \ \label{beta-weak}
\ee
while the strong disorder regime gives
\be
\log_2 \beta_- = -1 \ , \ \log_2 \beta_+ =  {4 \over q_c} -1 \ . \ \label{beta-strong}
\ee

The phenomenon of ESS reflects, in loose words, the fact that some fundamental statistical property of the self-similar cascade keeps holding in the crossover towards non-multifractal behavior, while a set of parameters (as the fragmentation parameter $a$ or the dimension $d$) evolves ``adiabatically".
Let us assume that the probability density of $\beta$ is fixed along the cascade, in the random $\beta$-model framework.
We are able to establish, in this way, two independent phenomenological pictures that support ESS.

{\it{Picture A: Target Space with Fractal Dimension}}.
We assume in this case that fragmentation, initially defined in $d$ dimensions, has been gradually modified to occur in a $d/ \alpha$-dimensional space. Considering that the probability density determining the relative number of boxes generated within coherence regions does not vary, the only way to preserve the statistics of $\beta$ is from a modification of the parameter $a$, as may be inferred from (\ref{bm1}). We have
\be
\beta_n={  N_{(n+1)}  \over  N_n  }  a^{-d} =
{  N_{(n+1)}  \over  N_n   } (a')^{-d / \alpha}  \label{bm4}
\ee
and, therefore, $a'=a^\alpha$. Performing the substitution $d \rightarrow d/ \alpha$ and $ a \rightarrow a^\alpha$ in the 
expression (\ref{expnt}), we obtain $\zeta(q) \rightarrow \zeta(q) / \alpha$ for strange attractors or in the localization problem, which implies ESS-I in these systems.

{\it{Picture B: Enhanced/Supressed Fragmentation}}.
We consider here the situation where fragmentation is enhanced or supressed. That is, the relative number of boxes generated within coherence regions gets a factor $c$. The fragmentation takes place in $d$-dimensional space. As in the previous picture, the only way to keep the probability density for $\beta$ fixed is from a modification of the parameter $a$. We have
\be
\beta_n={ {c N_{(n+1)} } \over {  N_n }  } a^{-d}=
{ { N_{(n+1)} } \over { N_n } } (a')^{-d} \ . \ \label{bm5}
\ee
Defining $\alpha$ from $c \equiv a^{d(1 - \alpha )}$, we get, from (\ref{bm5}), $a'=a^\alpha$. We will have now ESS-II, with the crossover functions given by $f_1(r)=r^{d(1-1/ \alpha)}$ and $f_2(r) = r^{1 / \alpha}$.

An arbitrary combination of pictures $A$ and $B$ is the most general situation, leading always to ESS-II.
It is interesting to note, in passing, that both of these pictures yield ESS-II for the turbulence cascade.

\section{Conclusion}

We studied infra-red effects in an exactly solvable two-dimensional metal-insulator transition modeled by random Dirac fermions, observing in a clear way the existence of ESS-I and ESS-II, in the strong and weak disorder regimes, respectively.
To our knowledge this is the first verification of ESS for a system that is not manifestly classical. A cascade description of the multifractal probability density profile was proposed along the lines of the random $\beta$-model, in a spirit similar to the usual applications performed in strange attractors and turbulence. Furthermore, we found that strange attractors can in fact show ESS, from a straightforward numerical analysis of the H\'enon and Lorenz dynamical systems. 

Two independent pictures, defined within the random $\beta$-model approach, were put forward as possible phenomenological descriptions of ESS. A very interesting (and challenging) problem is to check, then, if these pictures are somehow realized in the multifractal systems where ESS holds. We believe, however, that the next natural step of research concerns the issue of ESS in other metal-insulator transitions already known in condensed matter. It is worth mentioning that in the two-dimensional localization problem, $\tau(1)=\zeta(1)-2=0$ is an exact relation \footnote{The exact result $\zeta(1)=d$ in $d$-dimensional space follows from translation invariance of the box probabilities. In the language of the random $\beta$-model, we have, according to (\ref{expnt}), $\langle \beta \rangle = 1$.}, in perfect analogy with Kolmogorov's $4/5$ law of turbulence. Therefore, it is likely ESS can play for the subject of localization the same important role it has for turbulence, providing better evaluations for the multifractal exponents of structure functions.

\section{acknowledgements}

This work has been partially supported by FAPERJ.

\begin{figure}

\caption{}
ESS-II data for structure functions in the weak disorder regime ($q_c=10$), taking results for $q=3,6,9$.

\caption{}
ESS-I data for structure functions in the strong disorder regime ($q_c=1$), taking results for $q=-3,-6$.

\caption{}
ESS-II data for the H\'enon attractor, where structure functions of order $q=3,6$ are taken into account.

\caption{}
ESS-I data for the Lorenz attractor, where structure functions of order $q=3,8$ are taken into account.

\end{figure}

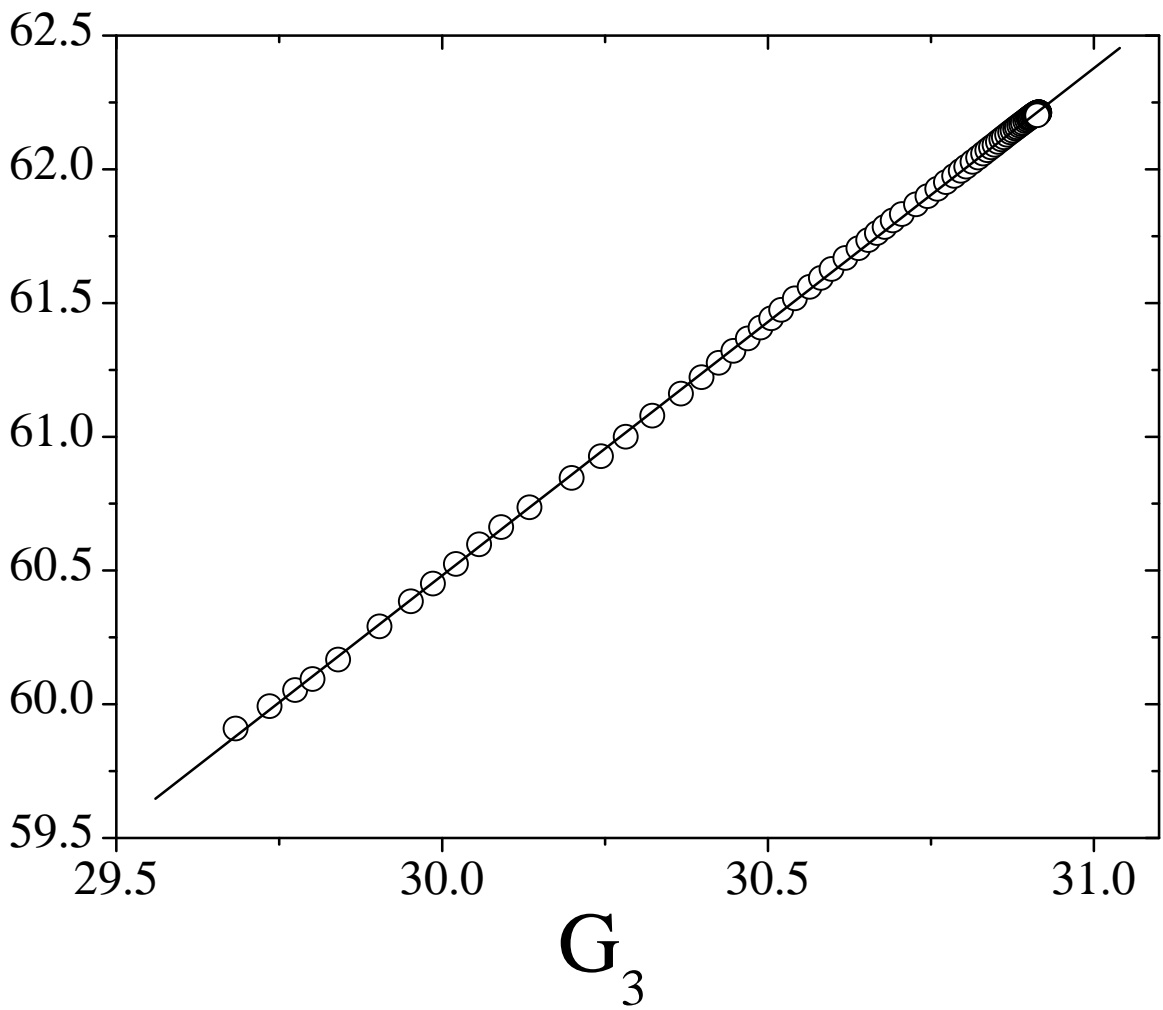
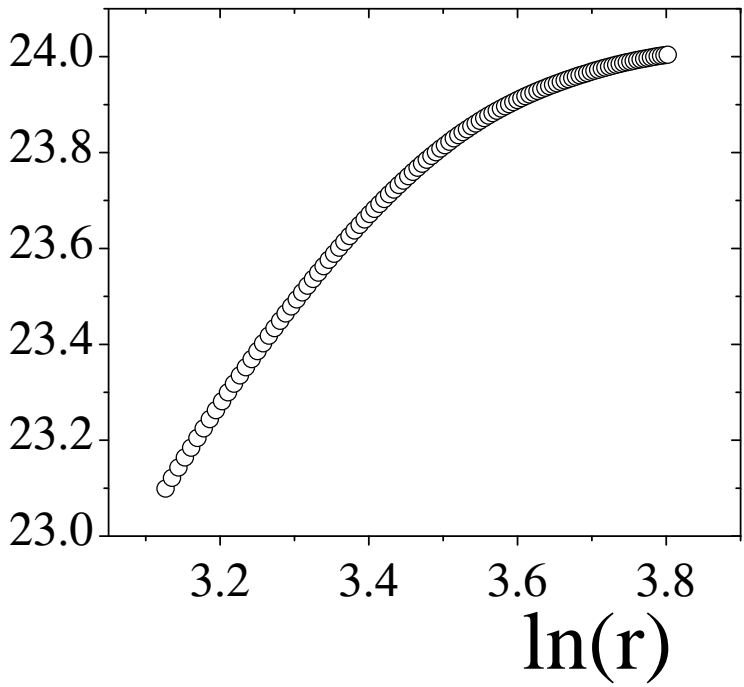
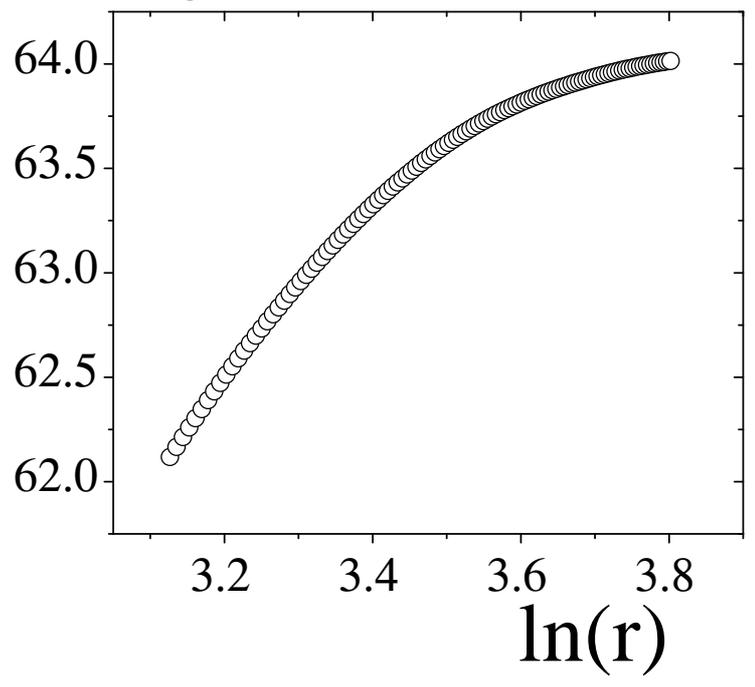
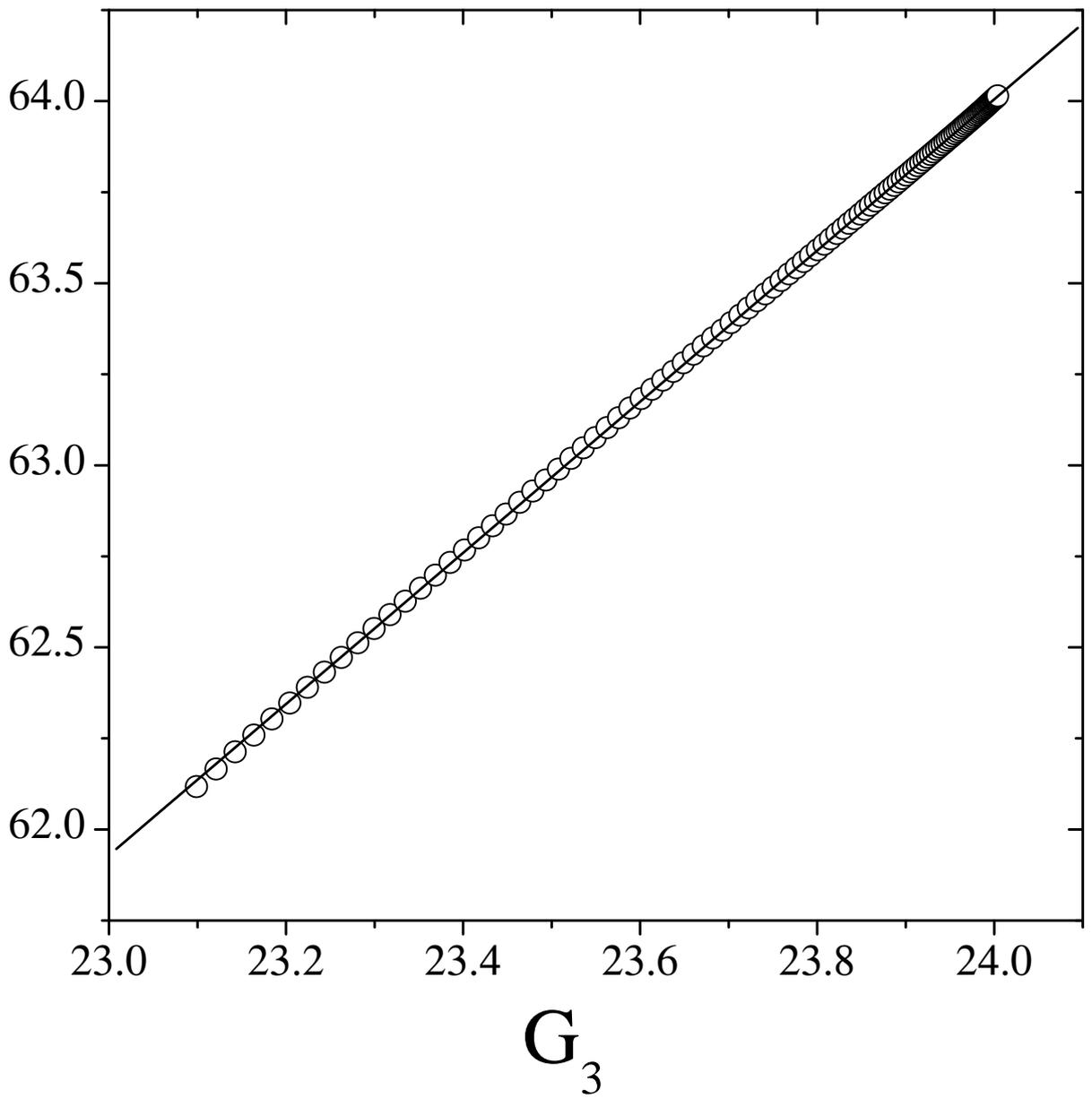
\end{document}